\documentclass{article}
  \usepackage{emulateapj}
  \usepackage{psfig}

\lefthead{Brandner et al.}
\righthead{Timescales of disk evolution and planet formation}

\begin{document}

\title{Timescales of disk evolution and planet formation:\\
HST, Adaptive Optics, and ISO observations of weak-line and post-T 
Tauri stars}\footnote{Based on 
observations with the NASA/ESA Hubble Space Telescope obtained at the Space 
Telescope Science Institute, which is operated by the Association of 
Universities for Research in Astronomy, Inc., under the NASA contract 
NAS5-26555, observations at the European Southern Observatory, La Silla,
(ESO Prop ID 58.E-0169), and observations with ESA's Infrared Space 
Observatory.}

\author{Wolfgang Brandner\altaffilmark{2,3,4}, 
Hans Zinnecker\altaffilmark{5},
Juan M.\ Alcal\'a\altaffilmark{6},
France Allard\altaffilmark{7},
Elvira Covino\altaffilmark{6},
Sabine Frink\altaffilmark{8},
Rainer K\"ohler\altaffilmark{5},
Michael Kunkel\altaffilmark{9},
Andrea Moneti\altaffilmark{10},
Andreas Schweitzer\altaffilmark{11,4}}

\affil{$^2$University of Hawaii, Institute for Astronomy, 2680 Woodlawn Dr., 
Honolulu, HI 96822, USA; brandner@ifa.hawaii.edu}
\affil{$^3$University of Illinois at Urbana-Champaign, Department of Astronomy,
Urbana, IL 61801, USA}
\affil{$^4$Visiting astronomer, Cerro Tololo Inter-American Observatory,
NOAO, which are operated by AURA, Inc., under contract with the NSF}
\affil{$^5$Astrophysikalisches Institut Potsdam, An der Sternwarte 16,
D-14482 Potsdam, Germany; koehler@aip.de,hzinnecker@aip.de}
\affil{$^6$Osservatorio Astronomico di Capodimonte Napoli, Italy;
jmae@cerere.na.astro.it,covino@astrna.na.astro.it}
\affil{$^7$CRAL, Ecole Normale Superieure, 46 Allee d'Italie, Lyon, F-69364 France, Cedex 07; fallard@cral.ens-lyon.fr}
\affil{$^8$Center for Astrophysics \& Space Sciences, University of California 
San Diego, La Jolla, CA 92037-0424, USA;
sabine@ucsd.edu}
\affil{$^9$Astronomisches Institut der Universit\"at W\"urzburg,
Am Hubland, D-97074 W\"urzburg, Germany; Michael.Kunkel@skf.com}
\affil{$^{10}$IEM-CSIC, Serrano 121, 28006 Madrid, Spain; amoneti@isis.iem.csic.es}
\affil{$^{11}$University of Georgia, Dept.\ of Physics \& Astronomy 
and  Center for Simulational Physics, Athens, GA 30602-2451, USA; andy@hobbes.physast.uga.edu}

\begin{abstract} 
We present high-spatial resolution HST and ground-based adaptive
optics observations, and high-sensitivity 
ISO (ISOCAM \& ISOPHOT) observations of a sample of X-ray selected
weak-line (WTTS) and post (PTTS) T\,Tauri stars located in the nearby 
Chamaeleon T and Scorpius-Centaurus OB associations.
HST/NICMOS and adaptive optics observations aimed at identifying substellar 
companions
(young brown dwarfs) at separations $\ge$ 30 A.U.\ from the primary
stars. No such objects were found within 300 A.U.\ of any of the
target stars, and a number of faint objects at larger separations 
can very likely be attributed to a population of field (background) stars. 
ISOCAM observations of 5 to 15 Myr old
WTTS and PTTS in ScoCen reveal infrared excesses which are clearly 
above photospheric levels, and which have a spectral index intermediate
between that of younger (1 to 5 Myr) T Tauri stars in Chamaeleon and that of
pure stellar photospheres.
The difference in the spectral index of the older PTTS in ScoCen compared 
to the younger classical and 
weak-line TTS in Cha can be attributed to a deficiency of smaller size 
(0.1 to 1\,$\mu$m) dust grains relative to larger size ($\approx$5\,$\mu$m)
dust grains in the disks of the PTTS. The lack of small dust grains is
either due to the environment (effect of nearby O stars and supernova 
explosions) or due to disk evolution. 
If the latter is the case, it would hint that circumstellar disks
start to get dust depleted at an age between 5 to 15 Myr.
Dust depletion is very likely related to the build-up of larger particles
(ultimately rocks and planetesimals) and thus an indicator for the onset of the
period of planet formation.

\end{abstract}


\section{Introduction}

In the solar system, the coplanarity of planetary orbits and their moons, and 
the preferentially prograde rotation direction led Kant (1755) to the 
suggestion that the solar system evolved out of a flattened, disk like 
structure (``Urnebel''). Kant also proposed that planetary systems similar to
the solar system might be common around other stars.
Two of the major observational breakthroughs in astronomy in the 1990s were the
direct imaging detection of circumstellar disks around young
stars in nearby starforming regions with the Hubble Space Telescope
(O'Dell et al.\ 1993; McCaughrean \& O'Dell 1996; Burrows et al.\ 1996), 
and the indirect detection of 
giant planets in close orbits around nearby stars 
(Mayor \& Queloz 1995; Marcy \& Butler 1996). 
These important observational findings provide
strong support for Kant's hypotheses.
Still uncertain, however, are the timescales of disk evolution
and the exact physical processes leading to the formation of
giant and terrestrial planets.

Giant planets in the solar system possess a core of higher density material 
surrounded by a shell of metallic hydrogen and an outer atmosphere.
According to one model, a higher density (rocky) core with a mass of
$\approx$10\,M$_\oplus$ has to form first, before noticeable amounts of nebular
gas can be accreted by the proto-giant planet. Simulations indicate
that at least 10$^6$ yr are necessary to form a 10 M$_\oplus$
rocky core (Lissauer 1987), and that another 10$^7$ yr are required
for the 10 M$_\oplus$ core to accrete 300 M$_\oplus$ of nebular
gas.  It is still unknown if massive circumstellar disks
can indeed survive for such an extended period. A second model,
recently reinstated by Boss (1998),
suggests that gravitational instability of a protoplanetary disk
leads directly to the formation of a giant gaseous protoplanet on time scales
as short as 10$^3$ yr. The rocky core then forms due to the settling of
dust grains initially acquired, and by further accretion of solid bodies
in the course of the following 10$^5$\,yr.
The difference in timescales for the formation of giant planets in the two 
models provides observational means to decide for or against
either model by studying the  circumstellar environment of stars with ages
$\le$ 15\,Myr.

Weak-line (WTTS) and the even more evolved post-T Tauri stars (PTTS) are 
prime targets since -- 
contrary to  the very young classical T\,Tauri stars -- they lack 
strong 1.3 mm dust continuum emission. This suggests that WTTS (and PTTS)
no longer possess massive cold circumstellar dust disks (Beckwith et al.\ 1990,
Henning et al.\ 1993). In WTTS and PTTS, the (dusty)
circumstellar matter was either already partially accreted onto the central 
star or redistributed to form planetesimals or -- via disk
fragmentation -- to form directly giant planets.

We devised a twofold strategy in order to study
disk evolution and the timescales of the formation of planetary
systems. First, taking advantage of the high sensitivity of the
Infrared Space Observatory (ISO, Kessler et al.\ 1996), 
we searched for evidence of circumstellar disks around the
presumably diskless WTTS and PTTS. Secondly, using high-spatial
resolution imaging, we aimed at directly detecting faint substellar
companions to the WTTS and PTTS.

Section 2 describes the sample selection.
In Section 3 we explain the observing strategies and give an overview of
 the data reduction and analysis.
Section 4 describes the attempt to detect substellar companions
using the NASA/ESA Hubble Space Telescope (HST) and ground-based adaptive 
optics (AO). The results of the ISO observations and their implications are 
discussed  in Section 5. An overall summary of the findings is presented in
Section 6.

\section{Sample Selection}

\subsection{Selection of star forming regions}

We selected the nearby ($\approx$150\,pc) Chamaeleon T association and the
Scorpius-Centaurus OB association for our study. Young low-mass stars
in OB assocations are of particular interest, as most stars -- including
our Sun (see Vanhalla 1998, and references therein) -- are believed to have 
originated in OB associations (e.g., Miller \& Scalo 1978).
In the course of photometric and spectroscopic follow-up
studies of ROSAT sources, an initial sample of about 150 young, X-ray active 
late-type stars
was identified in the Chamaeleon T association (Alcal\'a et al.\ 1995,
1997; Covino et al.\ 1997) and the Scorpius-Centaurus OB association
(Kunkel et al.\ 2000).

\subsection{Classification of evolutionary status}

Traditionally, ages for pre-main sequence stars are derived based
on their location above the main-sequence and by comparison with theoretical
evolutionary tracks and isochrones in an HR-diagram. This approach relies
on precise distance estimates in order to derive luminosities.
In ScoCen and Cha, however, parallax measurements are only available for a
handful of the X-ray active low-mass stars (see Frink et al.\ 1998;
Neuh\"auser \& Brandner 1998; 
Kunkel et al.\ 2000). Furthermore, the individual parallax measurements
indicate a relatively large spread in distance among various members of these 
associations. Hence it would be advantageous to determine the evolutionary 
status of the X-ray active low-mass stars using properties which are independent
of the distance, like, e.g., spectral features. 

In general, the X-ray active low-mass stars in Chamaeleon exhibit higher 
lithium equivalent widths than their counterparts in Scorpius-Centaurus.
As lithium is continuously destroyed near the bottom of the convections
zone, the photosheric lithium abundance decreases with increasing stellar age
(Herbig 1965; Bodenheimer 1965). Therefore the lower lithium abundance indicates
that the low-mass stellar population in ScoCen is older than in Chamealeon.

Mart\'{\i}n (1997, 1998) suggested a refined classification, 
based on H$\alpha$ and lithium equivalent widths as a function 
of spectral type to distinguish between CTTS, WTTS and PTTS.
In a diagnostic diagram, late-type PTTS occupy a region intermediate
between WTTS and stars in young open clusters like the Pleiades or
$\alpha$ Per with ages of 60 to 120 Myr. Unfortunately, the lithium
criterion can only be applied to stars of spectral type K0 and later,
and does not allow to discriminate pre-main-sequence from 
zero-age-main-sequence (ZAMS) stars of earlier spectral types. 
Thus the number of stars located in the PTTS region of the diagnostic
diagram represents only a lower limit to the total number of PTTS in 
a stellar population (Mart\'{\i}n 1997).
Therefore additional criteria like stellar kinematics
(e.g., Feigelson 1996; Frink et al.\ 1997, 1998) are needed in
order to distinguish between pre-main-sequence and ZAMS stars of
spectral type K0 and earlier.


In order to avoid a contamination of our sample by ZAMS field stars,
unclassified stars according to the diagnostics by Mart\'{\i}n (1997,
1998) were largely rejected.
In Chamaeleon, the majority of the young late-type stars are CTTS and WTTS
and have typical ages between 1\,Myr and 5\,Myr (Alcal\'a et al.\ 1995).
In the Scorpius-Centaurus OB association, a total of 76 TTS
were detected. 49 of them can be classified as PTTS, 20 as WTTS and 7 as CTTS.
42 X-ray active stars in ScoCen remain
unclassified. Typical ages for the WTTS and PTTS in ScoCen are of the
order of 5\,Myr to 15\,Myr (Kunkel et al.\ 2000).

The sample was narrowed down to about 50 WTTS and PTTS by 
selecting preferentially late-type stars with lithium equivalent
width $\ga$ 0.1 to 0.6 \AA.

\subsection{Preselection against binary systems}

As we aimed for the search for faint (substellar)
companions and the study of disk evolution, it is important to consider
the effect of binary stars.  In binary systems, the complex dynamics and
gravitational interactions between the individual components and
their circumstellar and circumbinary disks might influence the
evolution of circumstellar disks and aggravate or even
completely inhibit the formation of substellar companions
(Papaloizou \& Pringle 1977; Artymowicz \& Lubow 1994). Recent
observations of circumstellar disks in multiple systems support these
theoretical predictions (e.g.\ HK Tau, Stapelfeldt et al.\ 1998,
Koresko 1998; HV Tau, Monin \& Bouvier 2000).
All stars in the sample were therefore surveyed for visual and 
spectroscopic binary companions (Brandner et al.\ 1996; Covino et al.\ 1997;
K\"ohler et al.\ 2000; Kunkel et al.\ 2000). None of the detected companions
was faint enough to be classified as a substellar source.
Binary and multiple systems with separations less than 3$''$ (450 A.U.\ at
a distance of 150\,pc) were excluded from the final target list. 
Table \ref{tab3} gives an overview on the physical properties of the
stars which were observed with HST and ISO.
 
\section{Observations and data reduction} 

\subsection{HST/NICMOS observations}

\begin{figure*}[htb]
\centerline{\psfig{figure=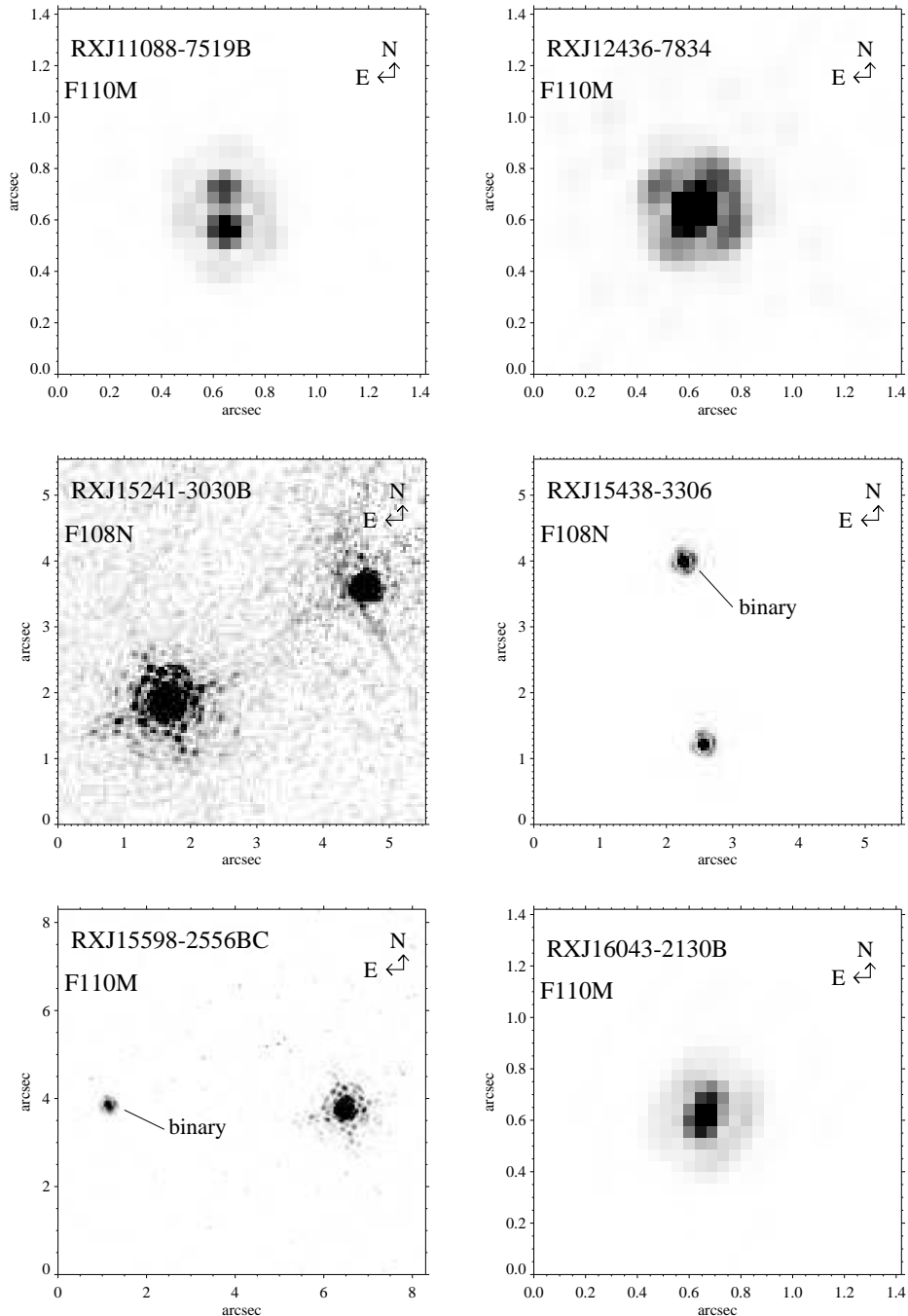,angle=0,width=13cm}}
\figcaption{Finding chart for binary and triple systems.
Two of the three wide binary systems in our survey turned out
to be hierarchical triple systems (RXJ 15438-3306 and RXJ 15598-2335BC).
North is up and east is to the left.
\label{fig1}}
\end{figure*}

\begin{figure*}[htb]
\centerline{\psfig{figure=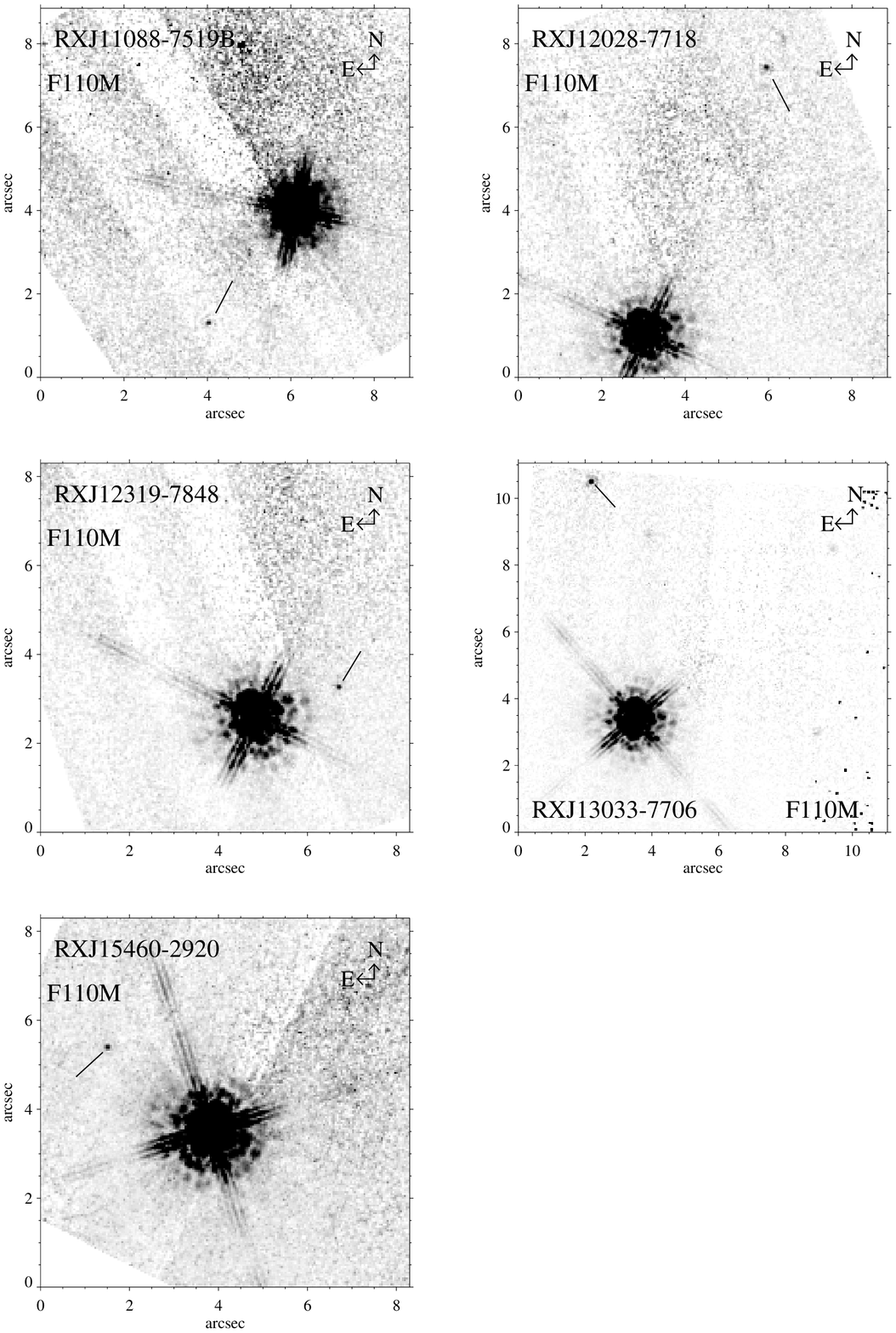,angle=0,width=13cm}}
\figcaption{Faint field (background?) sources.
North is up and east is to the left.
\label{fig2}}
\end{figure*}

\subsubsection{Observing Strategy}

Young substellar companions are still relatively hot, and are thus considerably
more luminous and much easier to spot than evolved (older and cooler) 
substellar companions (Brandner et  al.\ 1997; Malkov, Piskunov, \& Zinnecker 
1998). Maybe somewhat counterintuitive is the fact that
the best wavelength region to search for substellar companions
is the near-infrared between 1\,$\mu$m and 2\,$\mu$m.
The spectral energy distribution of late-type dwarfs and thus of young brown
dwarfs and giant planets is very peculiar (Allard \& Hauschildt 1995;
Burrows et al.\ 1997).
The molecular opacities which globally define
the continuum cause the spectral energy distribution to peak between
1\,$\mu$m and 2\,$\mu$m, almost independently of the effective temperature.
Using the HST/SYNPHOT simulator and ``Intermediate model spectra''
for brown dwarfs and low-mass stars from Allard et al.\
(1996), we simulated the expected brightness differences between
a substellar companion with an effective temperature of 900\,K, and its 
primary star with an effective temperature of 3800\,K and 2800\,K, 
respectively, when observed through various NICMOS filters (Table \ref{tab1}).
The narrow-band F108N filter and the intermediate band F165M filter 
provide the smallest brightness difference between the primary and a substellar
companion. For the HST observations we selected F108N because of its
shorter wavelength which yields a 35\% better spatial resolution compared to
F165M. In addition,  the narrowband F108N filter provides a
monochromatic, very well 
defined diffraction pattern around the science target, which is rather
independent of the color of the object. 
More recent evolutionary models for low-mass stars and ``NextGen model
spectra'' (Allard et al.\ 1997, Baraffe et al.\ 1998, Hauschildt et al.\ 
1999) largely confirm our choice of filter selection for the HST/NICMOS
observations.

In order to minimize the brightness difference between the primary and any 
possible substellar companion even further, we selected M-type WTTS and
PTTS because of their smaller intrinsic luminosity compared to earlier
spectral type WTTS and PTTS.
Simulations with the TinyTim PSF simulator (Krist \& Hook 1997) indicated
that it would be possible to detect an object 5$^{\rm m}$ fainter than
the primary at a separation of 0\farcs2. At a distance of 150\,pc, a
separation of 0\farcs2 corresponds to 30\,A.U., i.e.\ comparable to the
semimajor axis of Neptune's orbit.

\subsubsection{Observations}
The NICMOS camera aboard HST was installed in February 1997
(Thompson et al.\ 1998).
HST/NICMOS observations (see Table \ref{tab2}) of 24 M-type WTTS and PTTS 
in Chamaeleon and Scorpius-Centaurus were carried out between 1997 Aug 14 
and Sep 22 using NICMOS camera 1 (NIC1, 0\farcs043/pixel). The observations 
were obtained in the F108N filter using MULTIACCUM mode. 18 of the
targets were also observed in the F110M filter in order to obtain
additional color information. Exposure times
were between 2560s to 3072s in F108N, and between 64s to 192s in F110M. 

\subsubsection{Data reduction and analysis}
The data reduction was carried out using the IRAF/STSDAS package
CALNICA V3.1. For the data reduction, we replaced the model dark frames
(as provided by STScI) with darks derived from archived on-orbit measurements.
The photometric accuracy of the flux measurements for the faint sources
is limited by the uncertainties in the dark/bias subtraction, 
flatfield errors, and the resulting uncertainties in the local background. 
The overall photometric uncertainties are of the order of 0\fm05 for the
brighter objects, and 0\fm1 to 0\fm2 for the fainter sources. Figures
\ref{fig1} and \ref{fig2} show examples of the reduced frames.

As the NICMOS PSF can vary significantly from one HST orbit to the next,
the HST/NICMOS observations of all target stars were used to build up a 
library of HST/NIC1 PSFs in the F108N and F110M filters. For
each set of observations, we then identified the best matching
PSF in the library. Sub-pixel offsets between
target and PSF were computed by cross-correlating the individual frames.
The PSF was then Fourier-shifted, scaled, and subtracted. 
The resulting difference frames were searched for faint,
close companions. Figure \ref{fig4} shows the typical detection 
limit (solid line) in terms of brightness difference versus separation. 
It would have been possible to detect a 4\fm7 fainter
companion at a separation of 0\farcs2, and a 8$^{\rm m}$ fainter companion
at a separation of 1$''$. This detection limit is in good agreement with the 
detection limit derived from the simulations based on TinyTim PSFs.

Three of the ``single'' stars turned out to be close binaries, and
two of the three wide binaries turned out to be hierarchical triple
systems (Figure \ref{fig1}). A number of faint objects, which
were found close to some of the target stars, are shown in Figure \ref{fig2}.

\subsection{Adaptive Optics Observations}

The observing strategy for the adaptive optics observations with
ADONIS/SHARP at the ESO 3.6m telescope was similar to
that of the HST/NICMOS observations. Because of the need for a
star sufficiently bright for wavefront sensing, only bright G- and K-type
stars could be observed. The observations were carried out in the
broadband H and K filter, with the intent to gather additional
data with the circular variable filter (CVF) for good candidates 
for substellar companions.

In two observing nights in March 1997, a total of 17 WTTS and PTTS in 
Chamaeleon and Scorpius Centaurus were observed, and two 4$^{\rm m}$ to 
5$^{\rm m}$ fainter companions
were identified (RXJ 12253-7857 and RXJ 14150-7822, see
Brandner et al.\ 1997). Unfortunaly, unfavorable weather conditions
during the second half of the second observing night prevented any 
immediate follow-up using the CVF.

\subsection{ISOCAM \& ISOPHOT observations}

\begin{figure*}[htb]
\centerline{\psfig{figure=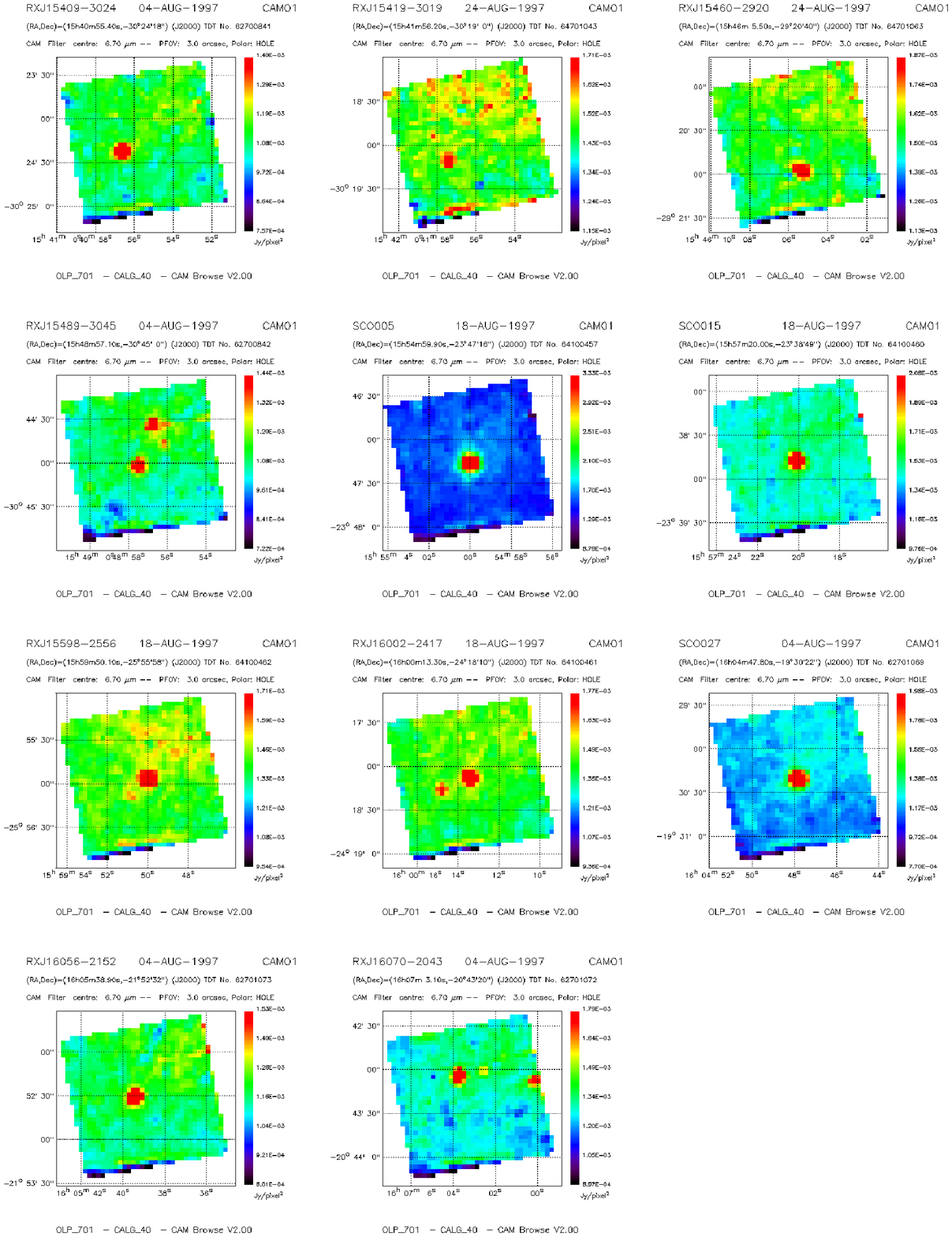,angle=0,width=15cm}}
\figcaption{ISOCAM 6.75\,$\mu$m observation of WTTS and PTTS in
Scorpius-Centaurs. In addition to the target stars, photospheric emission
from foreground stars has been detected in a number of cases. These
foreground stars are brighter than the target stars in the optical,
but considerably fainter in the infrared.
North is up and east is to the left.
\label{fig3}}
\end{figure*}

\subsubsection{Observing Strategy}

ISO offered the unique
opportunity to search for subtle infrared excesses from circumstellar
disks and substellar companions with unprecedented sensitivity.
As the temperature of circumstellar disks changes with
radial distance from the star, one can probe the radial surface density
and temperature profile of circumstellar disks by observing them at
a wide range of wavelengths. If for example in the course of disk evolution
the inner disk is depleted first, it should still be possible to
detect the cooler, outer parts of the disk at longer wavelengths. 
Consequently, the observing strategy was
to observe the sample stars both with ISOCAM (C\'esarsky et al.\ 1996)
at 6.7\,$\mu$m and 15\,$\mu$m, and with ISOPHOT (Lemke et al.\ 1996)
at 60\,$\mu$m and 90\,$\mu$m.

\subsubsection{Observations and Data Reduction}

The ISO observations with ISOCAM and ISOPHOT
were carried out between 1997 Aug 04 and 24.
Of the 30 WTTS and PTTS in the initial target list, only twelve
were actually observed with ISOPHOT. 
The ISOPHOT observations of twelve stars were
carried out as 3 $\times$ 3 maps using the PHT C100 detector in mode PHT22
and the C100\_60\_UM and C100\_90\_UM filters. NTTS155436-2313
was only observed with ISOPHOT. The other eleven
stars were also observed with ISOCAM.
The ISOCAM observations were obtained with a scale of
3$''$/pixel as 2 $\times$ 2 rasters using the LW2 (6.75\,$\mu$m) and
LW3 (15\,$\mu$m) filters. No color information could be obtained for
RXJ 15460-2920, as it was observed in LW2 only.
Total exposure times were of the order of 200s to 240s per filter
and object. 

The data reduction was carried out using the software packages CAM 
Interactive Analysis and
PHOT Interactive Analysis. The standard processing includes dark subtraction,
deglitching, transient correction, flatfielding and mosaicing. None of the
sources was detected at 60\,$\mu$m and 90\,$\mu$m down to a
limiting flux of $\approx$400\,mJy.

Figure \ref{fig3} shows the pipeline reduced ISOCAM data in the LW2 filter.
All eleven target stars (8 PTTS, 1 WTTS, 2 unclassified stars)
are clearly detected (Moneti et al.\ 1999). On several occasions
additional foreground (field) stars are also detected. Foreground stars
can easily be distinguished from
the WTTS and PTTS based on their LW3-LW2 colors (see below).
Aperture photometry was performed using IRAF and 4.5 pixel and
5.5 pixel apertures for LW2 and LW3, respectively.
Random uncertainties should
be of the order of a few percent, but systematic errors might be as large
as 20\% due to uncertainties in the transient correction and the
uncertain color terms. Table \ref{tab5} summarizes the results of the
ISOCAM observations.

\begin{figure*}[hb]
\centerline{\psfig{figure=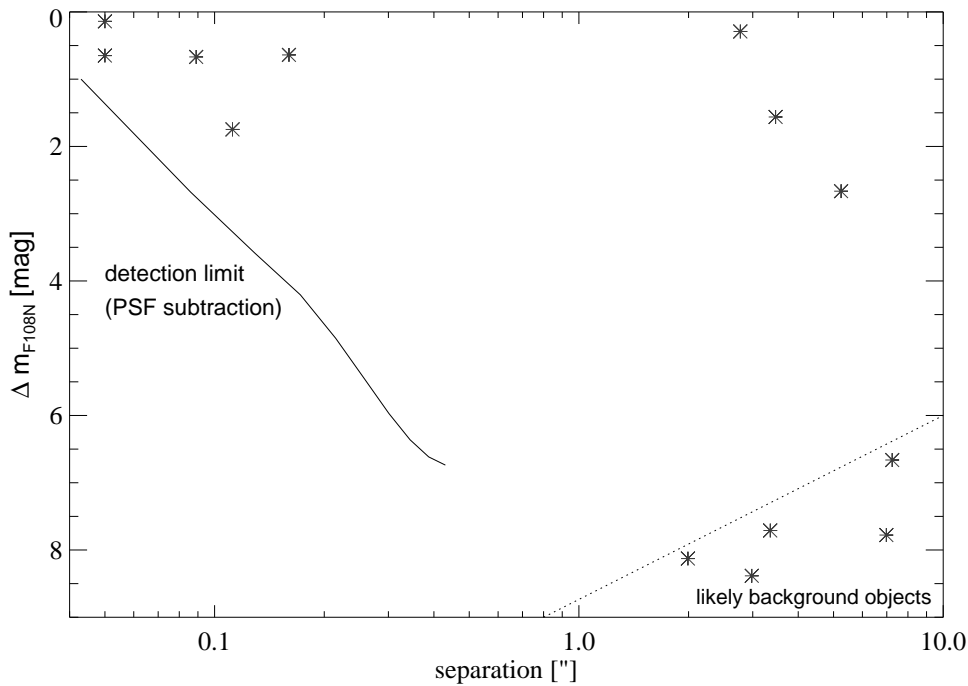,angle=90,width=14cm}}
\figcaption{Distribution of the brightness difference between the target
stars and their possible companions versus the angular separation.
The solid line indicates the detection limit after PSF subtraction.
It would have been possible to detect a 4\fm7 fainter companion
at a separation of 0\farcs2 and an 8$^{\rm m}$ fainter companion at a
separation of 1$''$, but none were found. The region preferentially occupied by
background objects is also shown (dotted line).
Note that the lack of nearly equal brightness binaries
with separation between 0\farcs2 and 2\farcs5 is due to the
preselection against binaries identified in the ground-based
speckle and direct imaging observations.
\label{fig4}}
\end{figure*}

\subsection{Ground-based NIR follow-up}

Ground-based near-infrared spectroscopy of the target stars and the faint
sources detected with HST/NICMOS and adaptive optics was attempted
in the course of two nights in April 1998 using the CTIO 4m Blanco telescope
and the near infrared spectrograph. Due to fog and clouds, no useful data
were obtained.

\section{Physical Companions and Background Sources}

\subsection{Faint (Substellar?) Sources}

The PSF subtraction of the HST/NICMOS data did not reveal any faint
objects which would qualify as a candidate for a substellar companion
within 2$''$ of the WTTS and PTTS.
A total of five point sources at projected separations $\ga$2$''$ and
 6 to 8 mag fainter than their
primary were detected in the 24 fields (Figure \ref{fig2}).
Photometric and astrometric measurements for the faint sources are summarized
in Table \ref{tab4}, and a plot of brightness difference between
the primary and a possible companion vs.\ separation is shown in
Figure \ref{fig4}.

\begin{figure*}[htb]
\centerline{\psfig{figure=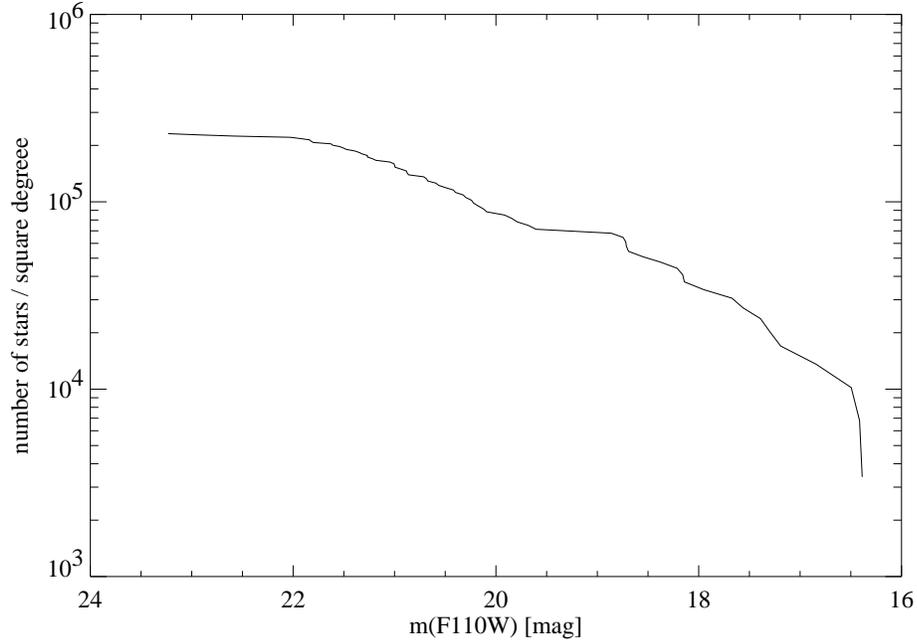,angle=90,width=13cm}}
\figcaption{Background star counts as measured on the NIC2 parallel
observations in the general direction of ScoCen (l$\approx$345$^\circ$,
b$\approx$+15$^\circ$) and scaled to 1 square degree. Based on these star
counts, we expect
on average 0.37 objects with 16\fm75$\le$m$_{\rm F110M}$ $\le$19\fm15
per NIC1 frame, or 4.0 background sources in total on the 11 NIC1 frames
towards ScoCen.
\label{fig5}}
\end{figure*}

\begin{figure*}[htb]
\centerline{\psfig{figure=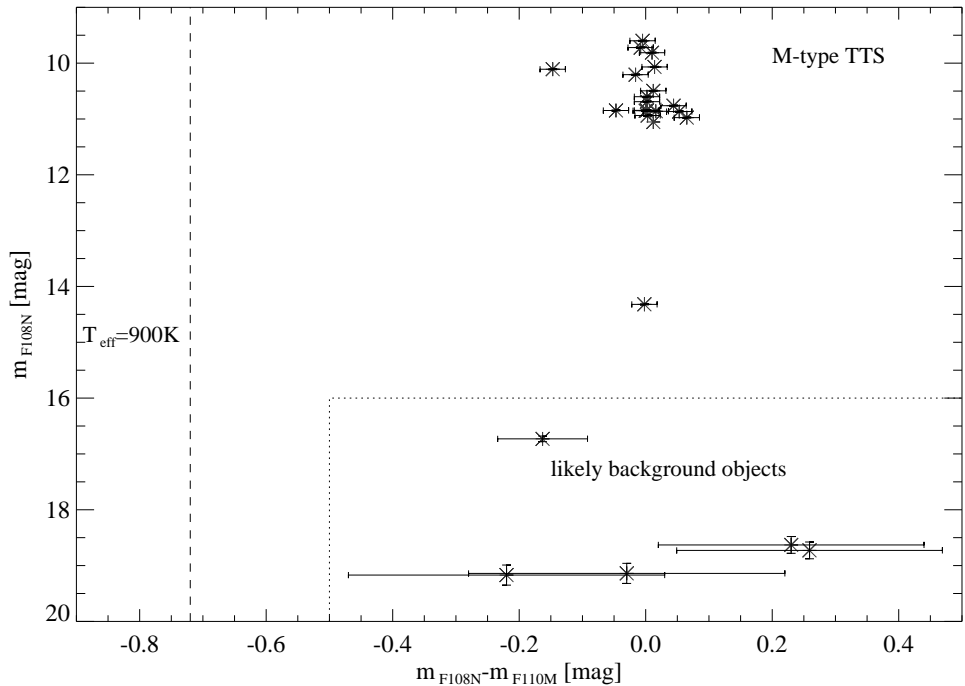,angle=90,width=14cm}}
\figcaption{Color magnitude diagram for target stars (WTTS and PTTS)
and faint sources. Most of the M-type TTS show m$_{\rm F108N}$-m$_{\rm F110M}$
colors close to 0 (with the exception of the PTTS RXJ 12046-7731).
Because of their faintness, the colors of the suspected background objects
have relatively large uncertainties, but all fall clearly outside
the region occupied by substellar objects. The dashed line
indicates the expected color for a brown dwarf with
T$_{\rm eff}$=900\,K.
\label{fig6}}
\end{figure*}

The probability $P(\Theta, \rm m)$ for an unrelated
source to be located within a certain angular distance $\Theta$ from
a particular target is given by

\begin{equation}
\label{eq:eps1} 
P(\Theta, \rm m) = 1 - e^{-\pi \rho (\rm m) \Theta^2}
\end{equation}

where $\rho (\rm m)$ is the cumulative surface density of background sources
down to a limiting magnitude m.
The probability for chance alignments increases
with increasing angular distance and decreasing brightness.
Therefore, in order to derive an estimate of $P(\Theta, \rm m)$, one
has to determine the local density of background sources first.

For the sources in Scorpius-Centaurus, parallel
NIC2 observations of adjacent fields were obtained in the F110W filter. 
The eleven NIC2 fields cover an area of about 4000 square arcsec.
68 sources with brightness values between 16\fm4 and 23\fm2 were detected.
Figure \ref{fig5} shows the cumulative brightness function $\rho(\rm m)$
scaled to an area of 1 square degree. The number counts in the
observed cumulative brightness function are in good 
agreement with the number counts expected according to the Galactic model
by Wainscoat et al.\ (1992).
No NIC2 parallel observations in F110W were obtained for the targets in
Chamaeleon, but a cumulative brightness function can again be derived from
the model by Wainscoat et al.\ (1992).

As the probability for physical association decreases with increasing
angular separation, our best candidate for a true substellar companion is 
the faint source $\approx$2$''$ north-west of RXJ12319-7848. According
to equation (\ref{eq:eps1}) the bare-bone probability for it being an unrelated
background source is only $\approx$5\%. In other words, equation (\ref{eq:eps1})
yields a 95\% probability for a physical association between
the RXJ12319-7848/c and RXJ12319-7848. Similar reasoning has in the
past frequently been cited as supporting evidence for the substellar
nature of faint objects near a brighter star (e.g., 0918-0023B, 
Jones et al.\ 1994; TMR-1C, Terebey et al.\ 1998; TWA-7B, Neuh\"auser 
et al.\ 2000a). 
In many cases, follow-up observations helped to establish the true nature
of the faint object, and the object turned out to be an
unrelated background star (e.g.,TMR-1C, Terebey et al.\ 2000;
TWA-7B, Neuh\"auser et al.\ 2000b) or a distant
galaxy (0918-0023B, Becklin et al.\ 1995).

It is therefore advisable to refrain from assigning probabilities
to individual sources. Instead one should compute the ensemble statistics.
While the probability for finding a background
object similar to RXJ12319-7848/c within 2$''$ of RXJ12319-7848 is only 5\%,
the probability for finding any such object within 2$''$
of one of the 24 HST/NICMOS targets is $\approx$70\%. It is
thus not unlikely that RXJ12319-7848/c is an unrelated background source.

Based on the observed cumulative brightness function for Scorpius-Centaurus,
one would expect to detect on average 4.0 sources with
F110W magnitudes between 16\fm75 and 19\fm15 on the eleven NIC1 frames
in the direction of Scorpius-Centaurus. One faint source was
actually detected, which is a 1.5 sigma deviation from the expected
average.
According to the model by Wainscoat et al.\ (1992) 
3.2 sources with J magnitudes between 16\fm75 and 19\fm15 would be
expected on average in a field equivalent in size to the area covered by the
13 NIC1 frames. Four faint sources were detected on the
NIC1 frames towards Chamaeleon, which is
in good agreement with the theoretical expectation.

The agreement between the m$_{\rm F108N}$ - m$_{\rm F110M}$ colors
of the faint objects and the M-type WTTS and PTTS (Figure \ref{fig6})
provides additional evidence that the faint objects are unrelated 
background sources and not substellar companions.

\subsection{Binary and multiple systems}

In addition to 21 stars which were assumed to be single stars, also three
wide binaries with separations $\ge$ 3\farcs5 were included in the sample.
Three of the ``single'' stars turned out to be close binaries, and
two of the three wide binaries turned out to be hierarchical triple
systems (see Figure \ref{fig1}.). RXJ 11088-7519 constitutes
an additional triple system, with RXJ 11088-7519a being a spectroscopic
binary (Covino et al.\ 1997).
The relatively high incidence of triple systems (see Figure \ref{fig1}) can 
be attributed to the selection effect in favor of multiple
systems in ROSAT selected samples of young X-ray active late-type stars 
(see Brandner et al.\ 1996; Kunkel et al.\ 2000). 

The adaptive optics observations led to the identification of two faint 
companions to X-ray active stars in Chamaeleon.
The relatively small lithium equivalent widths of RXJ 12253-7857
and RXJ 14150-7822 (see Alcal\'a et al.\ 1997) indicates that both
stars could be either pre-main-sequence stars with ages of up to 20\,Myr 
or ZAMS stars. In all cases the brightness of the secondary suggests that it is
a stellar companion (i.e., not substellar).

\subsection{Implications}

It seems likely that all five faint sources are
background objects, and not physically related to the PTTS and WTTS.
The lack of brown dwarf companions at separations larger than
30 A.U.\ (0\farcs2 at 150\,pc) indicates that wide brown dwarf companions are 
rare (less than 4\% of all cases).  
This is in good quantitative agreement with the
surveys for brown dwarf companions to white dwarfs (Probst 1983; 
Becklin \& Zuckerman 1988) and M-dwarfs (Oppenheimer 1999; 
Burgasser et al.\ 2000) in the solar neighborhood. It implies that
fragmentation of collapsing molecular cloud cores in general does not
produce extremely unequal mass pairs. Alternatively, the density of 
circumstellar disks at radii greater than 30 A.U.\ appears to be too 
small to allow for the formation of massive substellar companions.

\section{Evolved (Remnant) Circumstellar Disks}

\subsection{ISOCAM survey of low-mass star forming regions}

A large scale ISOCAM survey of low-mass star forming regions in the solar 
neighborhood was carried out by Nordh and collaborators (Nordh et al.\ 1996,
1998; Olofsson et al.\ 1999; Persi et al.\ 1999).  The first part
of Nordh's survey included the dark clouds Chamaeleon I, Ophiuchus,
Serpens, and Corona Australis and covered an area of 2.3 square degree.
A total of 402 sources in Oph, Cha, and CrA were detected both in LW2 and LW3
(Nordh et al.\ 1998). When placed on a $\log {\rm F}_{\nu}({\rm LW3})$
vs.\ $\log ({\rm F}_{\nu}({\rm LW3})/{\rm F}_{\nu}({\rm LW2}))$ magnitude-color
diagram, the sources cluster in two distinctive groups. 
As discussed by Nordh et al.\ (1996),
the two groups can be identified with disk-less main-sequence stars
(stellar photospheres), and pre-main sequence stars with circumstellar disks.

The expected value of
$\rm{R}$ = $\log \frac{F_\nu({\rm LW3})}{F_\nu({\rm LW2}) }$
for disk-less main-sequence
stars is approx.\ $-0.68$ and it is rather insensitive to the effective
temperature of the star. For T Tauri stars with circumstellar disks, R
can be derived from simple disk models.
Kenyon \& Hartmann (1995) computed the
spectral index

\begin{equation} \label{eq:eps2}
 \alpha_{12} = 
\frac{\log (\lambda_2 F_{\lambda_2}) - \log(\lambda_1 F_{\lambda_1})}{\log \lambda_2 - \log \lambda_1}
\end{equation}
for flat and flared disk models. According to their models, 
$\alpha = -0.7$ for a flared disk and $\alpha = -1.3$ for a flat disk at 
10\,$\mu$m. 
This corresonds to R=$-$0.1 and R=$+$0.1, respectively.
About 90\% of the 402 sources in the sample by Nordh et al.\ (1998) have
either R$<-$0.5 or R$>-$0.2, and are thus either (disk-less)
main-sequence stars or young pre-main sequence stars with circumstellar disks.
Only about 10\% of the sources fall in the transitional region
with $-0.5 \le \rm{R} \le-0.2$.

\subsection{ISOCAM mini-survey of Scorpius-Centaurus}

\begin{figure*}[htb]
  \psfig{figure=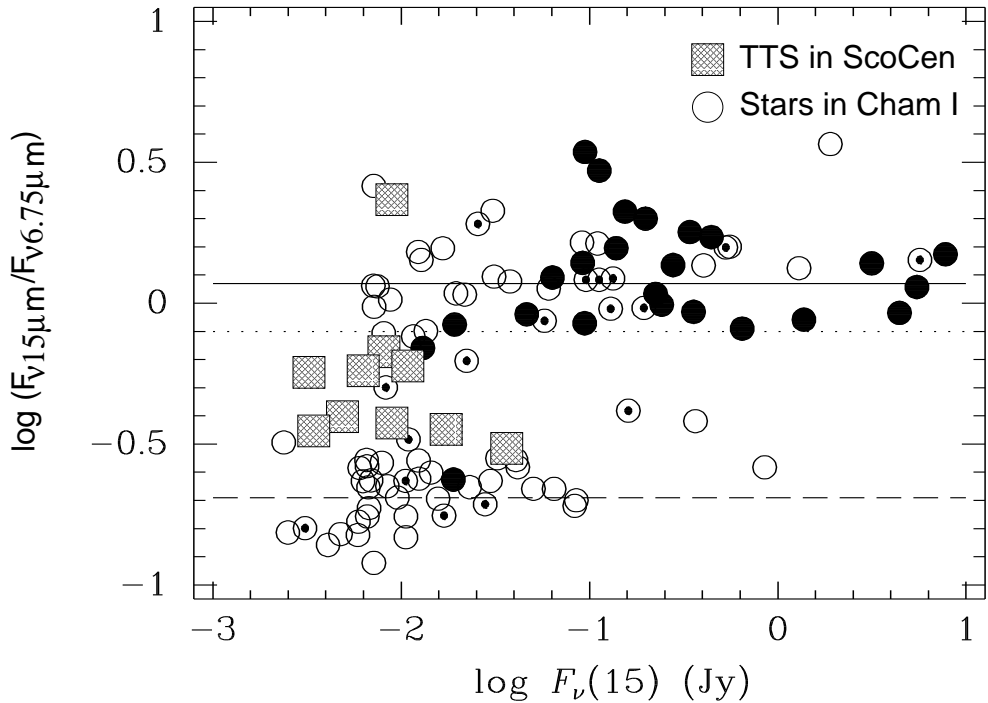,width=15.0cm,angle=270}
\figcaption{\label{isocmd}
(adapted from Nordh et al.\ 1996)
Color-magnitude diagram based on ISOCAM observations of
young stars in Chamaeleon (circles) and Scorpius-Centaurus (squares).
Black filled circles indicate
previously known YSOs and CTTSs, circles with a central dot
previously known WTTS, and open circles new sources
detected with ISO by Nordh et al. (1996). The dashed line indicates
the location of pure stellar photospheres, the dotted and solid
line the location of flat and flared circumstellar disks, respectively,
as predicted by the models from Kenyon \& Hartmann (1995).
WTTS and PTTS in ScoCen show a spectral index intermediate
between main-sequence stars and
CTTS and WTTS in Chamaeleon.
\label{fig7}}
\end{figure*}

Our ISOCAM mini-survey in Scorpius-Centarus covers only 10 stars
(1 WTTS, 7 PTTS, and 2 unclassified stars) 
which were observed both with LW2 and LW3. Two stars (1 WTTS, 1 PTTS)
have ISOPHOT observations, but no ISOCAM observations in LW3 (see Table
\ref{tab5}).
Figure \ref{fig7} shows one of the above mentioned
$\log {\rm F}_{\nu}({\rm LW3})$
vs.\ $\log ({\rm F}_{\nu}({\rm LW3})/{\rm F}_{\nu}({\rm LW2}))$ magnitude-color
diagrams for the sources detected towards Chamaeleon I by Nordh et al.\ (1996).
Overplotted are the sources from the present survey. 
Interestingly, the majority of the stars in our sample (seven out of ten)
have R in the range of $-$0.2 to $-$0.5, whereas the large majority
of the sources in Chamaeleon and other low-mass star forming regions
have R values outside this range. The values of R measured  for our sample of
PTTS are on average higher than the values for a pure stellar photosphere, 
but significantly less than the typical value measured for CTTS and WTTS
in low-mass star forming regions.

\subsection{Implications}

The non-detection with ISOPHOT at 60\,$\mu$m and 90\,$\mu$m confirms the
absence of cold, massive dust disks around the low-mass WTTS and PTTS in 
ScoCen. The ISOCAM observations, however, indicate the presence of 
circumstellar material (presumably in form of a circumstellar disk). 
The physical properties of the disks in ScoCen appear to
be different from the disks typically found around CTTS and
WTTS in low-mass star forming regions. The difference in R (or
spectral index $\alpha$) can be explained by differences in the global 
dust opacities at 6.5\,$\mu$m and 15\,$\mu$m in both types of disks. 

Suttner, Yorke \& Lin (1999) studied dust coagulation in the envelopes
around young stellar objects. They simulated the effect of grain
size on the specific extinction coefficient. Figure 1 in their paper indicates 
that a change in the average dust grain size from 0.1\,$\mu$m to 5\,$\mu$m
leads to an increase of the specific extinction coefficient 
$\kappa_{\rm i}^{\rm ext}$ at 6.5\,$\mu$m by a factor of $\approx$5, whereas 
$\kappa_{\rm i}^{\rm ext}$ at 15\,$\mu$m remains virtually unchanged. 
Such an increase in $\kappa_{\rm i}^{\rm ext}$(6.5\,$\mu$m) is
equivalent to an increase in the spectral index $\alpha$.
A ``flat'' spectral index $\alpha$ = 0, e.g., 
corresponds to R=$-$0.35. Thus any mechanism which preferentially removes or
destroys smaller grains can provide a viable explanation for the observed
differences in the spectral index of PTTS in ScoCen and WTTS and CTTS
in low-mass starforming regions such as Chamaeleon.

The higher ratio of PTTS to WTTS
in Scorpius-Centaurus compared to Chamaeleon suggests that the X-ray active
low-mass stars in Scorpius-Centaurus are on average older than their
counterparts in Chamaeleon.  Alcal\'a et al.\ (1997)
assign ages  less than 5 Myr to the majority of the X-ray active stars in
Chamaeleon, whereas Kunkel et al.\ (2000) find average ages in the
range of 5 to 15 Myr for the WTTS and PTTS in Scorpius-Centaurus.
Disk evolution and grain growths in circumstellar disks
would lead to a deficiency in smaller grains. Such a deficiency in smaller
grains should only last until a critical number of massive particles like rocks 
and planetesimals has formed. Collisions between individual planetesimals
will then replenish the disk with
smaller grains. During the initial build-up period, however, there should
be a phase in which a disk is almost devoid of smaller grains.

Apart from the age difference, the PTTS in ScoCen are located in an OB 
association,
and thus -- unlike the WTTS and CTTS in T-associations -- subject to the
effects of nearby O stars and supernova explosions. Both impacting
shock fronts (Dwek et al.\ 1996) and high energy photons from nearby O stars 
and occasional supernova explosions (Voit 1991) could change the
distribution of grain sizes. Overall, however, the
contribution of Lyman continuum radiation to dust destruction should be
negligible as only grains with sizes of $\la$0.001\,$\mu$m are affected
(see discussion in Richling \& Yorke 1997).

Thus disk evolution appears to be the more likely explanation for the
difference in spectral index between CTTS and WTTS in Chamaelon
and the PTTS in Scorpius-Centaurus.

\section{Summary}

High-spatial resolution HST and ground-based adaptive
optics observations, and high-sensitivity
ISO (ISOCAM \& ISOPHOT) observations of a sample of X-ray selected
weak-line (WTTS) and post (PTTS) T\,Tauri stars located in the nearby
Chamaeleon T and Scorpius-Centaurus OB associations were obtained and
analyzed. 

The HST/NICMOS and adaptive optics observations aimed at identifying 
substellar companions (young brown dwarfs) at separations $\ge$ 30 A.U.\ 
(0\farcs2 at 150\,pc) from the primary stars.
While the sample was preselected against binary stars with
0\farcs2$\le$sep.$\le$3$''$, we detected 5 binary stars with separations
$<$0\farcs2. The largest brightness difference between a primary and a
secondary in these close binaries is 1\fm75. The relatively
small brightness difference indicates that all five secondaries at
separations $<$0\farcs2 are low-mass stars with masses clearly above the
hydrogen burning limit. Even though it
would have been possible to detect a 4\fm7 fainter object at a
sepearation of 0\farcs2 from the primary star, or a
8$^{\rm m}$ fainter object at a separation $\ge$1$''$
from the primary star, no such objects were 
found within 300 A.U.\ (2$''$) of any of the target stars. 5 objects at
separations larger than 2$''$, and 6\,mag to 8\,mag fainter than the
target stars can very likely be attributed to a population of field
(background) stars. We conclude that the formation of massive
substellar companions at separations larger than 30 A.U.\ from the
primary star is very unlikely.

ISOCAM observations of
WTTS and PTTS in ScoCen reveal infrared excesses which are clearly
above photospheric levels, and which have a spectral index intermediate
between that of younger (1 to 5 Myr) T Tauri stars in Chamaeleon and that of
pure stellar photospheres.
The difference in the spectral index of the older WTTS and
PTTS in ScoCen compared to the younger classical and
weak-line TTS in Cha can be
understood in terms of disk evolution, and could hint that circumstellar disks
start to get dust depleted at an age of around 5 to 15 Myr.
Dust depletion is very likely related to the build-up of larger particles
(ultimately rocks and planetesimals) and thus an indicator for the onset of the
period of planet formation.

In evolved disks,
the presence of larger grains resulting from dust coagulation and the
gradual build-up of larger bodies should manifest itself both in the scattering
properties at longer wavelengths and in spectral features. In the near
future, using SIRTF/IRS and SOFIA, it will become possible to study the
mid- and far-infrared properties of the circumstellar disks around the PTTS
in ScoCen and other nearby star forming regions. Furthermore, high-sensitivity
CO surveys with the Atacama Large Millimeter Array will be able to
detect also the gaseous component of the disks around the PTTS and WTTS
in ScoCen and Chamaeleon. These surveys will provide the final answer to the
question whether massive gaseous disks can survive long enough to
allow for slow accretion of gas onto proto-giant planets over 
timescales of 10$^7$ yr, or if giant planets have to form on much shorter time 
scales by disk instabilities.

\acknowledgements
Support for this work was provided by NASA through grant number
GO-07412.01-96A from the Space Telescope Science Institute,
which is operated by the Association of Universities for Research in Astronomy,
Inc., under NASA contract NAS5-26555.
This publication makes use of data products from 2MASS, which is a
joint  project of UMass and IPAC/Caltech, funded by NASA and NSF.
 We would like to thank the program
coordinator Chris Ready and the contact scientist Keith Noll for their help in
implementing the HST observing program, and Dan Potter for providing the
numbers of the theoretical background source density.
WB acknowledges support by You-Hua Chu and Deborah L.\ Padgett, and by the
National Science Foundation. AS acknowledges support from the
Deutsche Forschungsgemeinschaft (DFG) under grant 1053/8-1 and NASA ATP
grant NAG 5-8425 to the University of Georgia.


\begin{table}[htb]
\caption{\label{tab1}
The expected brightness difference in various NIC1 filters between
an M-type primary
with an effective temperature of 2800\,K (M6) or 3800\,K (M0), respectively, and
a young brown dwarf with T$_{\rm eff}$=900K. 
F145M and F165M provide 
color information, albeit at worse spatial resolution than F108N.
For comparison, the expected brightness difference in the
WFPC2 F1042M filter is shown.}
\begin{tabular}{lcc}
Filter& $\Delta$m$_{\rm T2800K/T900K}$& $\Delta$m$_{\rm T3800K/T900K}$\\
& [mag]& [mag]\\ \hline
F095N    &     7.9           &          9.3 \\
F097N    &     7.8           &          9.2 \\
{\bf F108N } &{\bf 5.2}    &    {\bf  6.2} \\
F113N    &     8.9           &          9.9 \\
F164N    &     6.0           &          6.9 \\
F166N    &     5.5           &          6.5 \\
F187N    &     8.5           &         10.0  \\ \hline
F090M    &     7.6           &          9.1 \\
F110M    &     5.9           &          7.0 \\
{\bf F145M }  &{\bf 7.0}    &    {\bf  8.2} \\
{\bf F165M } &{\bf 5.1}    &    {\bf 6.1} \\
F170M    &     5.8           &          6.9 \\ \hline
F110W    &     5.6           &          6.7 \\
F140W    &     5.5           &          6.6 \\
F160W    &     5.6           &          6.7 \\ \hline \hline
WFPC2:   &                     &                \\
F1042M   &     6.1           &          7.4 \\  \hline
\end{tabular}
\end{table}

\begin{table}[htb]
\caption{Target List. Coordinates are presented in columns 2 and 3. Column
4 lists whether the object has been observed with HST or ISO. Column 5 list
aliases when available.\label{tab2}}
\begin{tabular}{llllcrll}
Target &$\alpha$(2000) &$\delta$(2000) &HST/ISO & Alias\\
&[hms]   &[$\circ$ $'$ $''$] & obs. &  \\ \hline
RXJ08480-7854$^1$  &08 47 57.21 &-78 54 54.0 &HST    & \\
RXJ09029-7759$^1$  &09 02 51.85 &-77 59 35.2 &HST    &\\
RXJ10053-7749$^1$  &10 05 20.54 &-77 48 42.8 &HST    & \\
RXJ11088-7519B$^1$ &11 08 52.92 &-75 19 03.1 &HST    & \\
RXJ11498-7850$^1$  &11 49 32.60 &-78 51 00.7 &HST    & \\
RXJ11585-7754B$^1$ &11 58 27.57 &-77 54 44.8 &HST    & \\
RXJ12028-7718$^1$  &12 02 55.25 &-77 18 37.7 &HST    & \\
RXJ12046-7731$^1$  &12 04 36.78 &-77 31 34.2 &HST    & \\
RXJ12077-7953$^1$  &12 07 48.96 &-79 52 42.3 &HST    & \\
RXJ12197-7403$^1$  &12 19 44.24 &-74 03 56.9 &HST    &  \\
RXJ12319-7848$^1$  &12 31 56.61 &-78 48 32.0 &HST    & \\
RXJ12436-7834$^1$  &12 43 37.21 &-78 34 07.4 &HST    & \\
RXJ13033-7706$^1$  &13 03 04.84 &-77 07 02.0 &HST    & \\
RXJ15241-3030B$^2$ &15 24 12.99 &-30 30 56.0 &HST    &  \\
RXJ15409-3024$^2$  &15 40 56.50 &-30 24 23.3 &HST, ISO &  \\
RXJ15419-3019$^2$  &15 41 57.49 &-30 19 04.5 &HST, ISO &  \\
RXJ15438-3306$^2$  &15 43 51.58 &-33 06 29.4 &HST    &  \\
RXJ15460-2920$^2$  &15 46 05.26 &-29 20 52.1 &HST, ISO &  \\
RXJ15489-3045$^2$  &15 48 57.30 &-30 45 02.4 &HST, ISO &  \\
RXJ15549-2347$^{2,3}$  &15 54 59.9  &-23 47 18   &ISO    & Sco005 \\
NTTS155421-2330$^3$&15 57 20.0  &-23 38 49   &ISO    & Sco015 \\
NTTS155436-2313$^3$&15 57 34.4  &-23 21 11   &ISO    & Sco017 \\
RXJ15598-2556$^2$  &15 59 50.05 &-25 55 57.9 &HST, ISO &  \\
RXJ16002-2417$^2$  &16 00 13.32 &-24 18 10.2 &HST, ISO &  \\
RXJ16043-2130B$^2$ &16 04 21.07 &-21 30 41.9 &HST    &  \\
RXJ16047-1930$^{2,3}$  &16 04 47.8  &-19 30 23   &ISO    & Sco027 \\
RXJ16056-2152$^2$  &16 05 39.40 &-21 52 34.0 &HST, ISO &  \\
RXJ16070-2043$^2$  &16 07 03.75 &-20 43 07.3 &HST, ISO &  \\ \hline
\end{tabular}

$^1$Alcal\'a et al.\ 1995, 1997;
$^2$Kunkel et al.\ 2000;
$^3$Walter et al.\ 1994
\end{table}

\begin{table}[htb]
\caption{Physical Parameters: Spectral type, Lithium equivalent width,
V magnitude and V-J color are from Alcal\'a et al.\ (1995, 1997),
Kunkel et al.\ (2000) and Walter et al.\ (1994), except where
mentioned ootherwise.\label{tab3}}
\begin{tabular}{lccrllrrr}
Target& TTS-class& SpT& EW(Li) &V  &V$-$J  & m$_{\rm F108N}$ & m$_{\rm F108N}$-m$_{\rm F110M}$ &F$_\nu$(F108N) \\
& & & [\AA] &[mag]  &[mag]  & [mag]& [mag] &[mJy]\\ \hline
RXJ08480-7854 &W&M2 &0.61 &13.19 &3.89 &9.60&$-$0.005 & 296.5 \\
RXJ09029-7759 &P&M3 &0.50 &13.99 &\nodata &10.50&0.012 & 129.8 \\
RXJ10053-7749 &W&M1 &0.57 &13.41 &3.59 &10.21 &$-$0.016 & 169.5 \\
RXJ11088-7519B&P&M3 &0.50 &14.78 &\nodata &10.97 &0.065 & 83.7 \\
RXJ11498-7850 &P&M1 &0.50 &14.37 &\nodata &9.81 &0.010 & 243.4 \\
RXJ11585-7754B&W&M3 &0.60 &14.29 &3.97 &10.69&0.002& 108.6 \\
RXJ12028-7718 &W&M0 &0.60 &14.38 &\nodata &10.85 &0.000 &93.5 \\
RXJ12046-7731 &P&M2 &0.47 &13.78 &\nodata &10.11 &$-$0.147& 185.3 \\
RXJ12077-7953 &W&M4 &0.60 &14.52 &\nodata &10.85 &0.002 & 94.9 \\
RXJ12197-7403 &W&M0 &0.56 &13.12 &\nodata &10.09 &\nodata & 189.5 \\
RXJ12319-7848 &W&M1 &0.60 &14.17 &3.90 &10.60 &0.002 & 117.8 \\
RXJ12436-7834 &W&M0 &0.70 &13.13 &3.75 &9.72 &$-$0.008 & 265.0 \\
RXJ13033-7706 &W&K7/M0&0.60&13.23 &\nodata &10.07 &0.01 & 192.5 \\
RXJ15241-3030B&P&M1 &0.38 &13.58 &\nodata &10.95 &\nodata & 85.8 \\
RXJ15409-3024 &P&M2 &0.11 &14.53 &3.89$^1$&11.02 &\nodata & 80.2 \\
RXJ15419-3019 &P&M4 &0.19 &16.06 &4.29$^1$&12.18 &\nodata & 27.6 \\
RXJ15438-3306S &?&M3 &0.04 & 14.86&4.20  &11.56 &\nodata & 48.6 \\
RXJ15438-3306N &\nodata&\nodata &\nodata &\nodata &\nodata  &11.85 &\nodata &37.2  \\
RXJ15460-2920 &P&M0 &0.42 &13.46 &\nodata &10.76 &0.044 &101.6 \\
RXJ15489-3045 &P&M2 &0.09 &15.28 &4.03$^1$&11.66 &\nodata & 44.5 \\
RXJ15549-2347  &?&G2 &0.28 &8.93 &1.36 &\nodata  &\nodata & \nodata \\
NTTS155421-2330&P&M0 &0.28 &12.78 &3.09 &\nodata  &\nodata & \nodata \\
NTTS155436-2313&W&M0 &0.58 &13.63 &3.77 &\nodata  &\nodata & \nodata \\
RXJ15598-2556A&P &M2 &0.31 &14.21 &3.61 &10.94 &0.003 & 86.4 \\
RXJ15598-2556BC &\nodata &\nodata &\nodata &\nodata &\nodata &14.32 &$-$0.002 & 3.8 \\
RXJ16002-2417 &P&M0 &0.47 &13.66 &3.20$^1$&10.87 &0.053 & 92.2 \\
RXJ16043-2130B&P&M2 &0.38 &15.06 &4.62 &10.85 &$-$0.047 & 93.9 \\
RXJ16047-1930 &?&K3 &0.32 &11.17 &2.25 &\nodata  &\nodata & \nodata \\
RXJ16056-2152 &P&M1 &0.50 &14.26 &\nodata &10.86 &0.016 & 92.7 \\
RXJ16070-2043 &W&M1 &0.55 &14.51 &\nodata &11.05 &0.012 &77.6 \\ \hline
\end{tabular}

$^1$J magnitudes based on 2MASS Second Incremental Release
data products.
\end{table}

\begin{table}[htb]
\caption{Separation, position angle, and photometric measurements
for components of multiple systems and faint field (background?)
sources\label{tab4}}
\begin{tabular}{lrrrrc}
Target& separation& PA & m$_{\rm F108N}$
          & m$_{\rm F108N}$-m$_{\rm F110M}$ &F$_\nu$(F108N)\\
& [$''$]& [$^\circ$] & [mag] & [mag] &[mJy]\\ \hline
RXJ11088-7519Ba&\nodata &\nodata &11.45 &0.065 & 53.7 \\
RXJ11088-7519Bb&0.160$\pm$0.010  &2.8$\pm$0.4 &12.08 &0.065 & 30.0 \\
RXJ12436-7834a&\nodata &\nodata &10.39 &$-$0.04: & 140.0 \\
RXJ12436-7834b&0.05:  &155.3$\pm$1.8 &10.53 &0.02:& 123.0 \\
RXJ15241-3030Ba&\nodata  &\nodata &10.94 &\nodata & 84.9 \\
RXJ15241-3030Bb&3.467  &300.5$\pm$0.1 &12.52 &\nodata & 19.7 \\ 
RXJ15438-3306Na&\nodata  &\nodata &12.03 &\nodata & 30.9 \\
RXJ15438-3306Nb&0.05:  &17.8:&12.68 &\nodata & 17.0 \\
RXJ15598-2556Ca&\nodata  &\nodata &14.53 &$-$0.03 & 3.0 \\
RXJ15598-2556Cb&0.112$\pm$0.005  &211.0$\pm$0.3 &16.28 &0.08 & 0.60 \\
RXJ16043-2130Ba$^1$&\nodata  &\nodata &11.38 &0.02 & 56.5 \\
RXJ16043-2130Bb$^1$&0.089$\pm$0.005  &342.5$\pm$0.2 &12.05 &0.02 & 30.5 \\ \hline
RXJ11088-7519B/c&3.352$\pm$0.030 &142.44$\pm$0.39 &19.16$\pm$0.18 &$-$0.22$\pm$0.25 & 0.04  \\
RXJ12028-7718/c&6.980$\pm$0.008 &335.69$\pm$0.01 &18.63$\pm$0.15 &0.23$\pm$0.21 & 0.07 \\
RXJ12319-7848/c&1.992$\pm$0.002  &290.83$\pm$0.14 &18.73$\pm$0.15 &0.26$\pm$0.21 & 0.07\\
RXJ13033-7706/c&7.243$\pm$0.002  &10.19$\pm$0.01 &16.73$\pm$0.05 &$-$0.16$\pm$0.07 & 0.42\\
RXJ15460-2920/c&2.928$\pm$0.021  &50.37$\pm$0.31 &19.15$\pm$0.18 &$-$0.03$\pm$0.25 & 0.05\\ \hline
\end{tabular}

$^1$Comparison to K\"ohler et al.\ (2000) indicates that
RXJ16043-2130B rotates counterclockwise with a rate of approx.\
5$^\circ$ yr$^{-1}$.
\end{table}

\begin{table}[htb]
\caption{Flux values derived from ISOCAM
observations for 8 PTTS, 2 WTTS, and 2 unclassified stars.
The unclassified star RXJ 15549-2347 (Sco005), which is the only 
G-type star in the sample, is also the only star with R$<-0.5$, and thus
very likely a disk-less main-sequence star. Two stars have R$>-0.2$.
RXJ 15598-2556 is a triple system which is unresolved by ISO.
RXJ 15419-3019 has a spectral type of M4. \label{tab5}}
\begin{tabular}{lrrrcc}
Target &F$_\nu$(6.75\,$\mu$m) & F$_\nu$(15\,$\mu$m) & R & TTS class&Alias\\
 &[mJy] & [mJy] & &  & \\ \hline
RXJ15409-3024  & 12.5  &  4.5 &$-$0.45  &P&\\
RXJ15419-3019  &  4.4 &11.7 &$+$0.42 &P& \\
RXJ15460-2920  & 11.3 &\nodata &\nodata &P& \\
RXJ15489-3045  &  6.9 & 3.8 &$-$0.26 &P&\\
RXJ15549-2347  &114.7 &32.8 &$-$0.54 &?& Sco005 \\
NTTS155421-2330& 24.4 & 9.6 &$-$0.40 &P &Sco015 \\
NTTS155436-2313& \nodata &\nodata &\nodata &W& Sco017 \\
RXJ15598-2556  & 14.5 &  9.6 &$-$0.18 &P& \\
RXJ16002-2417  & 14.4 & 9.1 &$-$0.20 & P&\\
RXJ16047-1930  & 49.2 & 16.9 &$-$0.46 &? &Sco027 \\
RXJ16056-2152  & 15.9 & 7.2 &$-$0.34 & P&\\
RXJ16070-2043  & 14.0 & 7.5 &$-$0.27 & W&\\ \hline
\end{tabular}
\end{table}


\begin{thebibliography}{}

\bibitem{} Alcal\'a, J.M., Krautter J., Schmitt, J. H. M. M., Covino E., Wichmann R., Mundt R.\ 1995, A\&AS, 114, 109 

\bibitem{} Alcal\'a, J.M., Krautter J., Covino E., Schmitt, J.H.M.M., Wichmann R., 1997, A\&A, 319, 184 

\bibitem{} Allard, F., Hauschildt, P.H.\ 1995, ApJ 445, 433

\bibitem{} Allard, F., Hauschildt, P.H., Alexander, D.R., Starrfield, S.\
1997, ARA\&A 35, 137

\bibitem{} Allard, F., Hauschildt, P.H., Baraffe, I., Chabrier, G.\ 1996,
ApJ 465, L123

\bibitem{} Artymowicz, P., Lubow, S.H.\ 1994, ApJ 421, 651

\bibitem{} Baraffe, I., Chabrier, G., Allard, F., Hauschildt, P.H.\ 1998,
A\&A 337, 403

\bibitem{} Becklin, E., Zuckerman, B.\ 1988, Nature 336, 656

\bibitem{} Becklin, E., Macintosh, B., Zuckerman, B.\ 1995, ApJ 449, L117

\bibitem{} Beckwith, S.V.W., Sargent, A.I., Chini, R.S., Guesten, R.\ 1990, AJ,
99, 924

\bibitem{} Bodenheimer, P.\ 1965, ApJ, 142, 451

\bibitem{} Boss, A.P.\ 1998, ApJ, 503, 923


\bibitem{} Brandner, W., Alcal\'a, J. M., Kunkel, M., Moneti, A., Zinnecker, H.\
1996, A\&A, 307, 121

\bibitem{} Brandner W., Zinnecker H., Allard, F. 1997, in 
``Brown Dwarfs and Extrasolar Planets'', eds. R. Rebolo et al., 
ASP Conf.\ Ser.\ 134, p.\ 288 

\bibitem{} Brandner, W., K\"ohler, R.\ 1998, ApJ 499, L79

\bibitem{} Burgasser, A.J., Kirkpatrick, D., Cutri, R.M., McCallon, H.,
Kopan, G.\ et al.\ 2000, ApJ Letters, in press

\bibitem{} Burrows, C.J., Stapelfeldt, K.R., Watson, A.M., Krist, J.E., Ballester, G.E.\ et al.\ 1996, ApJ 473, 437

\bibitem{} Burrows, A., Marley, M., Hubbard, W.B., Lunine, J.I.,
Guillot, T.\ et al.\ 1997, ApJ 491, 856

\bibitem{} C\'esarsky. C.J., Abergel, A., Agn\`ese, P.\ et al.\ 1996, A\&A 315,  L32

\bibitem{} Covino, E., Alcal\'a, J.M., Allain, S., Bouvier, J., Terranegra, L.,
Krautter, J.\ 1997, A\&A, 328, 187

\bibitem{} Dwek, E., Foster, S.M., Vancura, O.\ 1996, ApJ 457, 244

\bibitem{} Feigelson, E.D.\ 1996, ApJ 468, 306

\bibitem{} Frink, S., R\"oser, Neuh\"auser, R., Sterzik, M.F.\ 1997, A\&A
325, 613

\bibitem{} Frink, S., R\"oser, S., Alcal\'a, J.M., Covino, E., Brandner, W.\
1998, A\&A 338, 442

\bibitem{} Hauschildt, P.H., Allard, F., Ferguson, J., Baron, E.,
 Alexander, D.R.\ 1999, ApJ 525, 871

\bibitem{} Henning, T., Pfau, W., Zinnecker, H., Prusti, T.\ 1993, A\&A 276, 129

\bibitem{} Herbig, G.H.\ 1965, ApJ, 141, 588

\bibitem{} Jones, H.R.A., Miller, L., Glazebrook, K.\ 1994, MNRAS 270, L47

\bibitem{} Kant, I.\ 1755, Allgemeine Naturgeschichte und Theorie des Himmels,
Leipzig

\bibitem{} Kenyon, S.J., Hartmann, L.\ 1995, ApJS 101, 117

\bibitem{} Kessler, M.F., Steinz, J.A., Anderegg, M.E., Clavel, J., 
Drechsel, G.\ et al.\ 1996, A\&A 315, L27

\bibitem{} K\"ohler, R., Kunkel, M., Leinert, Ch., Zinnecker, H.\ 2000, A\&A, in
 press

\bibitem{} Koresko, C.D.\ 1998, ApJ 507, L145

\bibitem{} Krist, J., Hook, R.\ 1997, The Tiny Tim User's Handbook,
Version 4.4, Baltimore:STScI

\bibitem{} Kunkel, M., Brandner, W., Yorke, H.W., Zinnecker, H., Neuh\"auser, R., Schmitt, J.H.M.M., Mayor, M., Udry, S.\ 2000, ApJ Supp., submitted

\bibitem{} Lemke, D., Klaas, U., Abolins, J.\ et al.\ 1996, A\&A 315, L64

\bibitem{} Lissauer, J.J.\ 1987, Icarus, 69, 249

\bibitem{} Malkov, O., Piskunov, A., Zinnecker, H.\ 1998, A\&A 338, 452

\bibitem{} Marcy, G.W., Butler, R.P.\ 1996, ApJ, 464, L147

\bibitem{} Mart\'{\i}n, E.L., 1997, A\&A, 321, 492

\bibitem{} Mart\'{\i}n, E.L., 1998, AJ, 115, 351

\bibitem{} Mayor, M., Queloz, D.\ 1995, Nature, 378, 355

\bibitem{} McCaughrean, M.J., O'Dell, C.R.\ 1996, AJ, 111, 1977

\bibitem{} Miller, G.E., Scalo, J.M.\ 1978 PASP 90, 506

\bibitem{} Moneti, A., Zinnecker, H., Brandner, W., Wilking, B. 1999, 
in "Astrophysics with Infrared Surveys: A Prelude to SIRTF", ASP Conf.
Series 117, eds. M.D. Bicay, C.A. Beichman, R.M. Cutri, B.F. Madore, 
p.\ 355

\bibitem{} Monin, J.-L., Bouvier, J.\ 2000, A\&A 356, L75

\bibitem{} Neuh\"auser, R., Brandner, W.\ 1998, A\&A, 330, L29

\bibitem{} Neuh\"auser, R., Brandner, W., Eckart, A., Guenther, E., Alves,
J., Ott, Th., Huelamo, N., Fernandez, M.\ 2000a, A\&A, 354, L9

\bibitem{} Neuh\"auser, R.\ et al., 2000b, in prep.

\bibitem{} Nordh, L., Olofsson, G., Abergel, A., Andre, P., Blommaert, J.\
et al.\ 1996, A\&A Letters 315, 185

\bibitem{} Nordh, L., Olofsson, G., Bontemps, S., Huldtgren, M., Kaas, A.A.\
et al.\ 1998, in Star Formation with the Infrared Space Observatory, eds.\
Joao Yun \& Rene Liseau, ASP 132, p.\ 127

\bibitem{} O'Dell, C.R.O., Wen, Z., Hu, X.\ 1993, ApJ, 410, 686

\bibitem{} Olofsson, G., Huldtgren, M., Kaas, A.A., Bontemps, S., Nordh, L.\
et al.\ 1999, A\&A 350, 883

\bibitem{} Oppenheimer, B.R.\ 1999, PhD thesis, Caltech, Pasadena, CA

\bibitem{} Papaloizou, J., Pringle, J.E.\ 1977, MNRAS 181, 441

\bibitem{} Persi, P., Marenzi, A.R., Kaas, A.A., Olofsson, G., Nordh, L.,
Roth, M.\ 1999, AJ 117, 439

\bibitem{} Probst, R.\ 1983 ApJ 274, 237

\bibitem{} Richling, S., Yorke, H.W.\ 1997, A\&A 327, 217

\bibitem{} Stapelfeldt, K.R., Krist, J.E., Menard, F., Bouvier, J., Padgett,
D.L., Burrows, C.J.\ 1998, ApJ 502, L65

\bibitem{} Suttner, G., Yorke, H.W., Lin, D.N.C.\ 1999, ApJ 524, 857

\bibitem{} Terebey, S., van Buren, D., Padgett, D.L., Hancock, T., Brundage, 
M.\ 1998, ApJ 507, L71

\bibitem{} Terebey, S., van Buren, D., Matthews, K., Padgett, D.L.\ 2000,
AJ, in press (May issue)

\bibitem{} Thompson, R.I., Rieke, M., Schneider, G., Hines, D.C., Corbin, M.R.\
1998, ApJ 492, L95

\bibitem{} Vanhalla, H.A.T.\ 1998, Proc.\ Indian Acad.\ Sci.\ (Earth Planet
Sci.) 107, 391


\bibitem{} Voit, G.M.\ 1991, ApJ 379, 122

\bibitem{} Wainscoat, R.J., Cohen, M., Volk, K., Walker, H.J., Schwartz, D.E.\
1992, ApJS 83, 111

\bibitem{} Walter, F.M., Vrba, F.J., Mathieu, R.D., Brown, A., Myers, P.C.\
1994, AJ, 107, 692

\end{thebibliography}
\end{document}